# High-Ratio Compression for Machine-Generated Data

JIUJING ZHANG*, Guangzhou University & University of New South Wales, China & Australia
ZHITAO SHEN, Ant Group, China
SHIYU YANG†, Guangzhou University, China
LINGKAI MENG, Ant Group & Antai College of Economics and Management, Shanghai Jiao Tong University, China
CHUAN XIAO, Osaka University & Nagoya University, Japan
WEI JIA, Ant Group, China
YUE LI, Ant Group, China
QINHUI SUN, Ant Group, China
WENJIE ZHANG, University of New South Wales, Australia
XUEMIN LIN, Antai College of Economics and Management, Shanghai Jiao Tong University, China

Machine-generated data is rapidly growing and poses challenges for data-intensive systems, especially as the growth of data outpaces the growth of storage space. To cope with the storage issue, compression plays a critical role in storage engines, particularly for data-intensive applications, where a high compression ratio and efficient random access are essential. However, existing compression techniques tend to focus on general-purpose and data block approaches, but overlook the inherent structure of machine-generated data and hence result in low compression ratios or limited lookup efficiency. To address these limitations, we introduce the Pattern-Based Compression (*PBC*) algorithm, which specifically targets patterns in machine-generated data to achieve Pareto-optimality in most cases. Unlike traditional data block-based methods, *PBC* compresses data on a per-record basis, facilitating rapid random access. Our experimental evaluation demonstrates that *PBC*, on average, achieves a compression ratio twice as high as the state-of-the-art techniques while maintaining competitive compression and decompression speeds. We also integrate *PBC* to a production database system and achieve improvements on both comparison ratio and throughput.

## 1 INTRODUCTION

Machine-generated data [62], automatically generated by various computer applications and services, is produced in a relatively high volume. The International Data Corporation predicts that the global amount of data will reach $1.75 \times 10^{23}$ Gigabytes by 2025, with machine-generated data

---

*Work done while the author was an intern at Ant Group.
†Corresponding author.

Authors' addresses: Jiujing Zhang, Guangzhou University & University of New South Wales, China & Australia, jiujing.zhang@outlook.com; Zhitao Shen, Ant Group, China, zhitao.szt@antgroup.com; Shiyu Yang, Guangzhou University, China, syyang@gzhu.edu.cn; Lingkai Meng, Ant Group & Antai College of Economics and Management, Shanghai Jiao Tong University, China, mlk123@sjtu.edu.cn; Chuan Xiao, Osaka University & Nagoya University, Japan, chuanx@ist.osaka-u.ac.jp; Wei Jia, Ant Group, China, jw94525@antgroup.com; Yue Li, Ant Group, China, ly321766@antgroup.com; Qinhui Sun, Ant Group, China, qinhui.sunqinhui@antgroup.com; Wenjie Zhang, University of New South Wales, Australia, wenjie.zhang@unsw.edu.au; Xuemin Lin, Antai College of Economics and Management, Shanghai Jiao Tong University, China, xuemin.lin@gmail.com.

(e.g., data records generated by computer processes and logs produced by different equipment, but not including data generated by Generative AIs, which is more like the natural language) being a significant driver of this growth [47]. Consequently, data-intensive systems require efficient methods to compress machine-generated data, especially as the growth of data is outpacing the growth of storage space [41].

Storage cost typically accounts for a significant portion of the overall cost for data-intensive services. For instance, the TierBase (often abbreviated as TBase internally) from Ant Group, a Redis-compatible, durable, distributed, in-memory database service, occupies thousands of servers and hundreds of clusters to provide high throughput and low latency for a wide range of production use cases (e.g., caching workloads for business data). In many of these cases, memory is in high demand, determining the prime hardware cost, whereas CPU utilization usually remains relatively low. In view of this situation, data compression offers a feasible way to save on hardware costs.

Many popular database engines (e.g., RocksDB [18], LevelDB [23], etc.) have already conducted data compression by utilizing general-purpose compression methods (e.g., Gzip [31], Zstd [11]). However, when evaluating these methods on a large corpus of industrial data from our production workloads and attempting to improve storage usage, we observe two issues:

(i) Most stored data is machine-generated, and is usually considered unstructured or semi-structured data, which can be derived into common structures.

```c
struct trade { // Define a structure to store trade records
  char symbol[10]; // Stock symbol
  char side; // Buy or sell
  int quantity; // Number of shares
  double price; // Trade price
  time_t timestamp; // Trade time
};

void to_json(struct trade record, char* output) {
    sprintf(output, "{symbol: %s, side: %c, quantity: %d, price: %.2f, timestamp:
        %ld}", record.symbol, record.side, record.quantity, record.price,
        record.timestamp);
}
```

Take the above code as a simple example; it may convert financial trade records to JSON with the same template defined with the sprintf function. A simple example of the generated record is {"symbol": "IBM", "side": "B", "quantity": 100, "price": 50.25, "timestamp": 1639574096}. From this example, we observe that the template in the JSON record takes 66 bytes, and the values only take 22 bytes. We can save significant storage space if we transform these generated records by only storing the valid values. In a set of data records with multiple complex patterns, if we can cluster data with the same pattern and only store each pattern once, storage space can be significantly saved.

(ii) In many applications, user requests for data access are not sequential. As a result, database systems often require random point access to data. For example, key-value storage systems like Redis can store the trade records above as values and enable efficient random retrieval and update operations. Unfortunately, it is challenging for general-purpose compression methods to compress each data record individually with a considerable compression ratio. Most of them adopt a block-wise compression strategy to compress the data, which leads to inefficient random access.

The first observation has been overlooked in previous work, and to the best of our knowledge, there is no existing research that leverages the inherent structural information of machine-generated data to enable more effective compression. As for the compression algorithms designed for specific data types, despite achieving better compression ratio than general-purpose methods, most of them are limited to specific data types or require strong assumptions about data structure. These data-type-specific algorithms can be divided into two categories: (1) Compression methods that rely on user-provided regular expressions or other structural knowledge to describe the data. For example, most JSON compression methods are based on the structure of JSON files. Log compression methods, such as Logzip [37] and LogReducer [58], require user-provided regular expressions or external log parsers about the uncompressed log to guide the compression. (2) Compression methods that can automatically discover the structure information of data. For instance, PIDS [32] can automatically extract sub-attributes from relational string attributes in columnar stores. By treating them as individual units to store and encode, PIDS can serve as a compression method that outperforms Snappy [24] and Gzip [31] in compression ratio. However, it can only handle data from a single source or with the same structure (e.g., the same column in a columnar database) and does not work with data having multiple structures.

However, machine-generated data originate from disparate sources with different formats, meaning that database engines typically have no prior knowledge (e.g., generation rules or data distribution) about the data. Moreover, even for data from the same tenant or namespace, complex business requirements often result in data with multiple and diverse structures, rendering methods like PIDS unable to parse the data structure.

The second observation, also mentioned by FSST [6], is that when compressing data records in data blocks, we need to decompress the entire data block first and then lookup the queried data record, which hinders fast random access to data. Therefore, compressing machine-generated data for data engines with high compression ratios and considerable decompression (and lookup) speeds remains challenging.

Motivated by these observations, we propose the Pattern-Based Compression (*PBC*) algorithm, which utilizes the implicit structural characteristics of machine-generated data by extracting common subsequences. Furthermore, unlike the block compression, the proposed *PBC* algorithm compresses data records individually, naturally supporting fast random access to data records. We also propose an efficient encoding length-based clustering framework that can automatically discover patterns from data collections without any prior knowledge. The proposed *PBC* algorithm achieves state-of-the-art compression ratios, potentially helping databases save over half of storage costs. We provide a theoretical analysis of the proposed algorithm, indicating the equivalence between minimal encoding length merging, which we use for clustering, and the minimal entropy clustering problem. As such, we show that the proposed clustering criterion guides compression towards minimizing entropy.

**Contributions.** The contributions are summarized as follows.

- To the best of our knowledge, this is the first systematic work to compress data records individually by reducing the redundancy in common subsequences.
- We propose the concept of encoding length and a corresponding optimization problem to model the compression problem. Based on this, we design efficient algorithms to cluster data with similar structures and automatically discover patterns.
- Extensive experiments are conducted on a variety of datasets. The experimental results demonstrate that *PBC*, the proposed algorithm, significantly outperforms state-of-the-art compression methods in compression ratio, as well as provides a competitive compression and decompression speed. The results also demonstrates that *PBC* achieves Pareto-optimality in many cases.

- Compared to block compression methods, *PBC* outperforms Zstd by up to two orders of magnitude in lookup speed and achieves state-of-the-art compression ratios. We also show that, even when compared to JSON/log-specialized compression techniques, *PBC* provides a comparable compression ratio and outperforms them in compression and decompression speeds.
- We integrate *PBC* into a production database system and evaluate it using real production data. The results from real business data confirm that *PBC* outperforms existing compression methods in both compression ratio and throughput.

**Organization.** The rest of the paper is organized as follows. We survey related works in Section 2. In Section 3, we provide a solution overview, including pattern extraction, compression, and decompression. In Section 4, we formalize the minimal encoding length clustering problem and present algorithms for extracting patterns from samples, with further optimizations presented in Section 5. In Section 6, we model the data compression from the perspective of entropy and examine the equivalence between the minimal encoding length clustering problem and the entropy minimization problem. We present experimental results in Section 7. Section 8 concludes this paper.

## 2 RELATED WORK

Owing to the escalating importance of storage space, numerous storage systems have adopted a variety of techniques to compress data records. Next, we distinguish between one-size-fits-all compression and compression methods specific to certain data types.

### 2.1 One-Size-Fits-All Compression Techniques

The most popular compression algorithms are one-size-fits-all, having no preference for data distribution and compressing data without any prior knowledge.

**LZ (Lempel–Ziv) family.** The majority of these algorithms belong to the LZ family, e.g., LZ77 [67] and its variants [11, 24, 31, 61, 68]. They are adopted to compress data records and save storage costs in modern NoSQL databases. For example, Apache Hadoop [51] uses the LZ4 [64] algorithm for fast compression, which is considered the best lightweight compression method [6, 52]. LevelDB [23] uses Snappy [24] to compress data, which is a fast data compression and decompression library based on LZ77. Similar to LZ4, it aims for extremely high compression performance rather than maximum compression. Snappy is widely used in Google systems like Bigtable [9], MapReduce [15], and Google's internal RPC systems.

AWS Redshift [48] and Facebook RocksDB [18] include support for field compression using Zstandard (Zstd) [11]. Zstd was designed to provide fast decompression speed and a high compression ratio comparable to the classical DEFLATE algorithm [60], which is the basis of original ZIP and Gzip. Zstd combines the dictionary-matching technique from LZ77 but using a large search window with efficient entropy coding techniques (Huffman coding [26], arithmetic coding [42], and Asymmetric Numeral Systems (ANS) [16]). Zstd also offers a special mode with offline dictionary training to improve the compression ratio on short data. Due to its advanced performance (faster than any other currently available algorithm with similar or better compression ratio), Zstd reaches the current Pareto frontier and is considered one of the most suitable compression methods for modern database systems [63].

**Grammar-based compression.** Grammar-based compression builds a context-free grammar (CFG) for the data to be compressed. Although the problem is known to be NP-hard, a few methods (e.g., Sequitur [40], Re-Pair [34], and GLZA [12]) were proposed to approximate it. However, grammar construction is quite expensive, and the size of grammar can be large and complex. Recent work improves Re-Pair by using AVX512 to speed up decompression [46], but it still remains costly in time [6].

**Tree-indexed Compressed Strings.** Modern DBMSs rely on in-memory search trees to achieve high throughput and low latency. Some of them (e.g., tries and radix trees [4, 35, 65]) store keys vertically and provide compression. They can deliver extremely fast query speed but with limited compression ratio, even when cooperating with a dictionary-based compressor [66].

## 2.2 Compression for Specific Data Types

Some specific types of data, such as machine-generated logs and attributes in columnar databases, are increasingly produced and stored in storage systems. Since the scale of these kinds of data is tremendous, compression is widely used to save storage costs.

**Compression for log data.** A series of works [10, 36, 37, 58, 59] attempt to compress log data, with the current popular approach being to capture the log structure using log parsers. Logzip [37] parses templates on log samples and extracts all templates in original log files through an iterative matching and extraction process. It then uses LZMA [61] or Gzip [31] to compress template IDs and variables separately. Based on the observation that numerical values account for a large percentage of space and Logzip [37] does not work well on it, LogReducer [58] is built upon Logzip and features specific encoding techniques for timestamps and numerical variables, which allows it to achieve a better compression ratio. However, due to its dependence on log parsers, it cannot compress other types of data.

**Compression for JSON data.** The prevalent approach to JSON-specific compression is to utilize JSON-compatible serialization specifications, some of which are designed for space efficiency, including BSON [8] (used in MongoDB [38]), CBOR [7], MessagePack [20] (used in Redis [45]), and JSONB (JSON Tiles) [17]. Viotti and Kinderkhedia conducted a comprehensive survey [55] and benchmarked [53] these specifications regarding space efficiency. They further proposed a serialization specification, *JSON BinPack*, alongside a reproducible benchmark study [54]. This study demonstrated that *JSON BinPack* outperforms 12 alternative binary serialization formats in all considered cases in terms of space efficiency. Another practical study from Lucidchart [50] suggests that *Amazon Ion* [1] is one of the most compact binary serialization alternatives for JSON.

**Compression for columns in columnar databases.** Columnar databases rely on data compression to save storage space and improve bandwidth utilization. In columnar databases, attributes usually have a similar structure. A series of methods were proposed [5, 22, 32, 39] to achieve better compression by utilizing this similarity. PIDS [32] can extract sub-attributes from relational string attributes in columnar stores. It mines common patterns in string attributes and uses them to split attributes. By encoding sub-attributes individually, PIDS can serve as a compression method that outperforms Snappy and Gzip in compression ratio. However, it can only handle data from a single source or with a similar structure (e.g., from the same column of a columnar database), making it unable to process the majority of machine-generated data due to their multiple structures.

Many columnar stores, such as Parquet [56], ORC [2], Carbondata [19], and DuckDB [43], allow for lightweight encoding methods (e.g., hard-coded encoding selection [2, 19]) over traditional byte-oriented compression algorithms like Snappy and Gzip. FSST [6] is a state-of-the-art lightweight compression method that mainly targets string columns. FSST is the first algorithm to directly support random access while offering a competitive compression ratio. Unfortunately, lightweight encoding methods only lead to a sub-optimal compression ratio in practice, e.g., the compression ratio of FSST on XML and JSON files can be 2 to 2.5 times worse than LZ4.

Typical column compression focuses on individual columns, often neglecting vital connections between columns or tuples. Semantic compression identifies these relationships and stores only the necessary columns/tuples for table reconstruction. Various methods, such as clustering [29, 30], regression trees [3], entropy compression [44], Bayesian networks [14, 21], and auto-encoders [27], have been employed to capture these relationships. Semantic compression, while similar to our work in identifying relationships, operates on different data units. It is designed to work solely on tokenized data in table format, which distinguishes it from *PBC* in terms of complexity and clustering criteria.

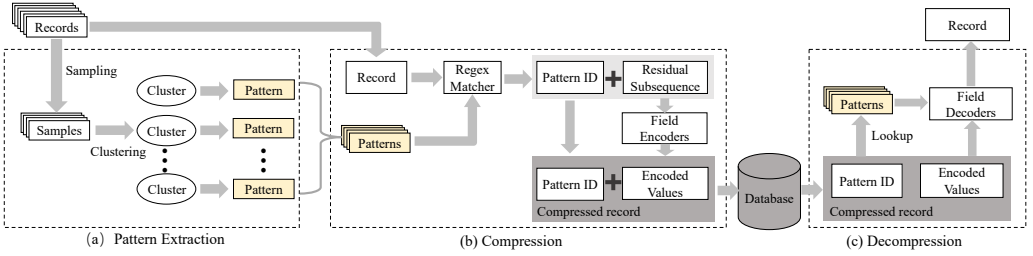

Fig. 1. Overview of the *PBC* framework.

## 3 SOLUTION OVERVIEW

### 3.1 Preliminaries

For simplicity, we assume data record is represented by string. In rest of this section, we use record and string interchangeably. Given a string $s$, let $len(s)$ denote its length, $s[i]$ denote the $i$-th element in $s$, and $s[i:j]$ denote the substring from $s[i]$ to $s[j]$, $1 \leq i \leq j \leq len(s)$. A sequence $s' = s[i_1], s[i_2], \ldots, s[i_m]$ is called a subsequence of $s$, where $1 \leq i_1 \leq i_2 \leq \cdots \leq i_m \leq len(s)$ and $len(s') = m$. A sequence $cs$ is called a common subsequence of a string set $S = \{s_1, s_2, \ldots, s_n\}$ if $cs$ is a subsequence of string $s_i, \forall s_i \in S$. We denote the set of all common subsequences of a string set $S$ as $CS(S)$.

### 3.2 Solution Framework

The idea of our solution is to reduce storage redundancy by encoding the duplicated subsequences among a data cluster. The compression consists of two phases: (1) the pattern extraction phase, which clusters the data records and extracts patterns from the clusters, and (2) the pattern-based compression phase, which compresses the original data using the extracted patterns. The decompression involves piecing together the original data from the pattern and the residual part of the data record (i.e., the part not covered by the extracted patterns). Figure 1 shows the overview of solution framework.

**Pattern Extraction.** Figure 1(a) illustrates the process of clustering records and extracting patterns using sampled data. The goal of this phase is to capture the structures of machine-generated data and mine potential patterns from the samples. Records are divided into clusters and patterns are extracted from them. A dictionary is then built to map each pattern to an integer code. Note that this phase is performed offline.

**Pattern-based Compression.** As shown in Figure 1(b), each record entry is matched to multiple patterns (described as regular expressions). If a record entry matches more than one patterns, the longest one is selected. To accelerate the matching speed, we adopt Hyperscan [57], a state-of-the-art regex matcher, for matching regular expressions. Once a pattern is determined for a record entry, its residual subsequence is extracted, and then the fields in the residual subsequence are encoded, according to the rules in Table 1.

Consider the instance of a record "foobar" and two patterns, "*ob*" and "*ooba*". Each pattern is converted into a regular expression, e.g., "*ob*" becomes "[.*]ob[.*]". A regex matcher is then employed to identify the pattern that matches the record. Here, both patterns match the record. We select the longest one (i.e., "*ooba*") to achieve a better compression ratio. Once the pattern is chosen, the residual subsequences (i.e., ["f", "r"]) can be extracted, as depicted in Figure 2. The record entry is restructured into a shorter concatenated code consisting of the pattern ID and the encoded values from the residual subsequence. If a record cannot match any pattern, it is treated as an outlier and saved in its raw form. In practice, when the number of outliers reaches a fixed percentage (e.g., due to data updates), the pattern extraction phase is restarted, in order to find new patterns to cover these outliers.

Table 1. Field encoders for residual subsequences.

| Encoder | Description |
|---|---|
| $CHAR(n)$ | Fixed length characters. $n$ indicates the length of substrings. |
| $VARCHAR$ | Variable length characters. Unlike $CHAR$, $VARCHAR$ holds 1 or 2 bytes header for the character length information. |
| $INT(n, m)$ | If all the substrings in the same field are fixed length digits. We can use integer to represent the digits. n indicates the length of digits and m indicates the bytes of converted integer. |
| $VARINT$ | For the digits without fixed length, we can use variable length unsigned integer encoder to encode numbers for space saving. |
| Further compression | To further reduce the storage redundancy, it is optional to process the residual subsequences with line-wise encoders (compress on records) or block-wise encoders. |

**Example 1.** Figure 2 illustrates an example of patterns, residual subsequences, and corresponding field encoders. Given a cluster with four records, a pattern for this cluster is a common subsequence of records with encoders for each field. For a cluster $c_i$, we denote its pattern as $Pat(c_i) = \{p_i, L\}$, where $p_i$ is a common subsequence of $c_i$ with a wildcard $*$ to represent the fields. The remaining subsequences in those records are residual subsequences. Each residual subsequence can be divided into a number of field values. Substrings that can be matched by the same wildcard are called a field, and values in the same field share the same encoder. In $Pat(c_i) = \{p_i, L\}$, $L$ is a list of encoders for each field. In this case, '*<$INT(2, 1)$>', '*<$INT(6, 2)$>', and '*<$VARCHAR$>' are the encoders of the fields. '*<$INT(2, 1)$>' and '*<$INT(6, 2)$>' represent wildcards that allow matching 2 and 6 digits, encoded with int8 and int16, respectively. '*<$VARCHAR$>' means a wildcard that allows matching characters with arbitrary length. In Figure 2, residual subsequences can be extracted using the pattern. The encoding method for each field in the residual subsequences is determined by the pattern, helping save storage space. For example, in field 0, all the residual substrings are two-digit numbers, which can be saved as a 1-byte int8 value instead of two-digit characters.

**Decompression.** As shown in Figure 1(c), the compressed record consists of a pattern ID and the corresponding number of encoded field values. Each pattern ID is translated via a lookup in the pattern dictionary to find the corresponding common subsequence and field decoders. Similarly, each value can be decoded by the field decoder. Then, we can concatenate the common subsequence and the decoded values to obtain the original record.

### 3.3 Applicability Scope

*PBC* is designed to compress data records that share a set of common subsequences, which is described as follows.

Given a set of strings, if we can identify a partition scheme that divides the set into $k$ clusters such that a common subsequence (denoted by $cs_i$, where $len(cs_i) > 0, \forall i \in [1, k]$) can be found in each cluster $C_i$, then we say that this string set shares a set of common subsequences.

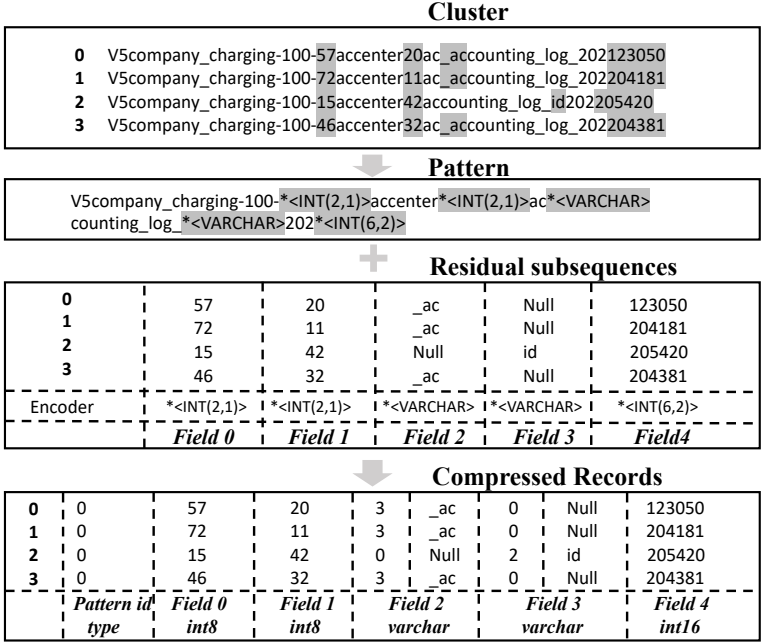

Fig. 2. Example of patterns, residual subsequences, and compressed records.

Many data types display this characteristic, and the compression ratio is influenced by the proportion of common subsequences. According to Wikipedia [62], machine-generated data is defined as data automatically produced by computer programs, applications, and services. This includes web server logs, call detail records, financial instrument trades, and other serialized data objects with various customized serialization rules (e.g., the majority of our proprietary data). The nature of being automatically generated by computer programs means that most machine-generated data records intrinsically share a set of common subsequences (though outliers may exist).

## 4 RECORD CLUSTERING AND PATTERN EXTRACTION

In this section, we formulate the problem of record clustering and pattern extraction, and then introduce the corresponding algorithms.

### 4.1 Problem Formulation

The primary algorithmic challenge within the proposed solution framework is to partition records into distinct clusters, enabling the discovery of optimal patterns that maximize raw data compression. To address this challenge, we formulate the clustering process as an optimization problem.

In practice, we not only build a pattern dictionary, but also often encode residual subsequences to reduce storage redundancy rather than storing the plain text directly. Specifically, we employ an encoding function $f : \{\Sigma \to \mathbb{R}^*\}$ to represent the method used to encode residual subsequences. Given the encoding function $f$ and a residual subsequence $r_1$, the encoding length under $f$ is $f(r_1)$.

First, we define the encoding length, which is used to quantify the compressed size of residual subsequences. We denote the combination of field encoders as an encoding function for residual subsequences.

*Definition 1.* **Encoding Length.** Given an encoding function $f : \{\Sigma \to \mathbb{R}^*\}$, and a set of strings $S = s_1, s_2, ..., s_m$ with a pattern $p$, the encoding length of $S$ under pattern $p$ and encoding function $f$ is $EL(S, p, f) = \sum_{i=1}^{m} f(r_i)$, where $r_i$ is the residual subsequence of $s_i$.

To denote the optimal encoding length that can minimize the final encoded data size, we define the minimal encoding length, optimal pattern and optimal encoding function:

*Definition 2.* **Minimal Encoding Length, Optimal Pattern, and Optimal Encoding Function.** Given a finite set of encoding functions $F = \{f_1, f_2, ..., f_{|F|}\}$, and a set of strings $S = s_1, s_2, ..., s_m$ with a pattern $Pat(S) = p$, the encoding length $EL(S, p, f)$ is minimal if there is no other $p', p' \neq p$ or $f', f' \neq f$ that satisfy $EL(S, p', f') < EL(S, p, f)$. We call the pattern $p$ optimal if there is no other $p', p' \neq p$ that satisfies $EL(S, p', f) < EL(S, p, f)$. We call the encoding function $f$ optimal if there is no other $f', f' \neq f$ that satisfies $EL(S, p, f') < EL(S, p, f)$. We denote the minimal encoding length as $EL_{min}(S)$.

Next, we formulate an optimization problem that minimizes the total size of the encoding length by finding the optimal clustering.

**PROBLEM 1** (MINIMAL ENCODING LENGTH CLUSTERING). *Given an encoding function set $F = \{f_1, f_2, ..., f_{|F|}\}$, a set of strings $S$, and a number constraint $k$, the Minimal Encoding Length Clustering problem aims to find an optimal clustering of $S \to \{S_1, S_2, ..., S_k\}$ and corresponding optimal patterns $P = \{p_1, p_2, ..., p_k\}$ such that $\sum_{i=0}^{k} EL(S_i, p_i, f)$ is minimal.*

**NP-hardness.** The minimal encoding length clustering problem is essentially a hierarchical clustering with an encoding length-based metric defined in Definition 1, and thus is NP-hard due to the NP-hardness of the hierarchical clustering [13].

## 4.2 Clustering Algorithm

To solve Problem 1 (*Minimal Encoding Length Clustering*), we adopt a greedy strategy similar to the agglomerative clustering, as shown in Figure 3: Initially, we treat each record as a cluster. In each iteration, we merge the two closest clusters, until the number of clusters reaches a given threshold $k$. Here, 'the two closest clusters' means the two clusters that lead to the minimal encoding length. To measure the closeness of the two clusters, we propose the notion of encoding length increment.

*Definition 3.* **Encoding Length Increment.** Given an encoding function set $F = \{f_1, f_2, ..., f_{|F|}\}$, two clusters $c_i$ and $c_j$, and the corresponding optimal patterns $p_i$ and $p_j$, the encoding length increment of merge $c_i$ and $c_j$ is

$$ELI(c_i, c_j) = EL_{min}(c_i \cup c_j) - EL_{min}(c_i) - EL_{min}(c_j). \tag{1}$$

Further, we formalize a problem that to find 'the two closest clusters'.

**PROBLEM 2** (MINIMAL ENCODING LENGTH MERGING). *Given an encoding function set $F = \{f_1, f_2, ..., f_{|F|}\}$, $N$ clusters $c_1, c_2, ..., c_N$, we call the merging of $c_i$ and $c_j$, $i \neq j$ and $i, j \in [1, N]$ the minimal encoding length merging, if the encoding length increment of merging $c_i$ and $c_j$ is minimal among all combinations of cluster pairs.*

A direct solution to Problem 2 (*Minimal Encoding Length Merging*) is to enumerate all possible results of merging every combination of cluster pairs and then merge the two clusters that lead to the minimal encoding length increment. We propose a dynamic programming algorithm to compute the encoding length increment of two given clusters.

**Example 2.** Figure 4 shows an example of the minimal encoding length increment computation of merging '$C_x = ab3*<VARCHAR>2$' and '$C_y = ab*<VARCHAR>12$'. We create a dynamic programming table with $5 \times 5$ entries. After filling them in a left-to-right, top-to-bottom order, the value of the lower-right entry is the minimal encoding length (arrows indicate where the final state is transferred from). We use two examples (Ex.A & Ex.B) to explain how we compute the encoding length and how the state transition works.

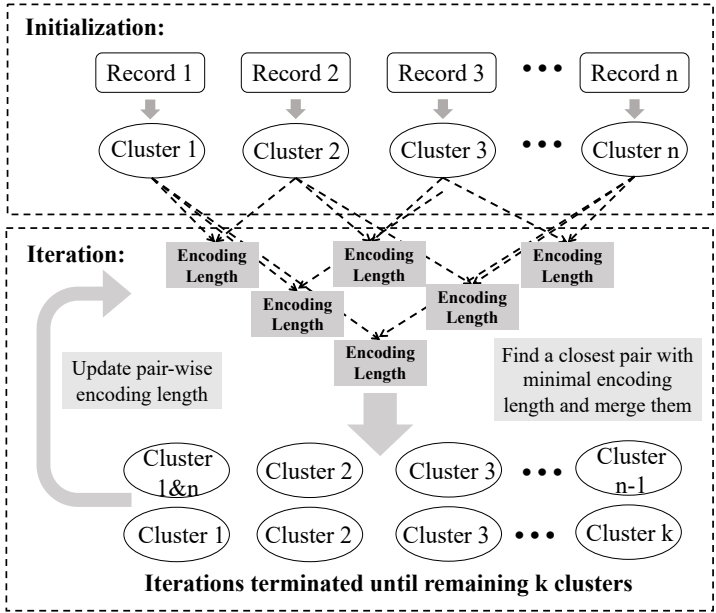

Fig. 3. Clustering framework.

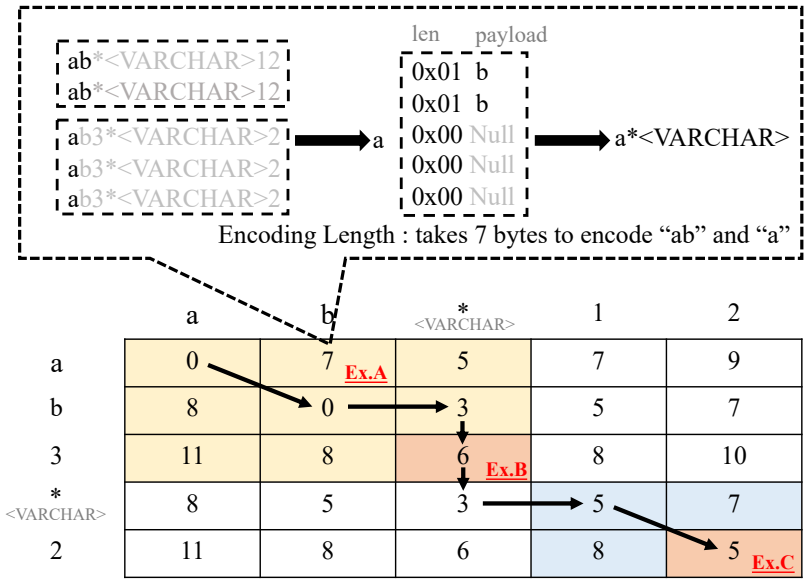

Fig. 4. Example of minimal encoding length increment merging.

**Ex.A**: The value of this entry (with value 7) represents the minimal encoding length increment of strings in the first '$a$' of $C_x$ : {$ab3*<VARCHAR>2$} and the first '$ab$' of $C_y$ : {$ab*<VARCHAR>12$}. In the in-lined example, we explain how we calculate encoding length increment with the VARCHAR encoder: After merging $C_x$ and $C_y$ into $C'$ : {$a*<VARCHAR>$}, the encoding length increment is the extra space required to store the increase of residual subsequences. In this case, the increment includes two '$b$' strings and corresponding wildcards. Thus, two '$b$'s take 2 bytes each for a 1-byte length and 1-byte payload, while three '$Null$'s each take 1 byte for zero-length strings. Therefore, the encoding length after merging the two clusters is 7.

**Ex.B**: We explain the origin of the value of the entry in Ex.B. To compute the value, we need to traverse all 8 entries from its upper-left (marked in yellow), then calculate the encoding length based on the state of these entries and choose the minimal possible result. The reason we need to go through all previous entries is to enumerate all the substrings for a possible wildcard replacement for each encoder.

**Algorithm Description.** Given an encoding function set $F$, two clusters $C_x, C_y$ with the optimal patterns $Pat(C_x) = \{cs_x \in CS(C_x), L_x\}, Pat(C_y) = \{cs_y \in CS(C_y), L_y\}$, this algorithm computes the minimal encoding length increment of the $C_x \cup C_y$ merging. Suppose $len(cs_x) = n$ and $len(cs_y) = m$, then the dynamic programming table contains $n \times m$ entries. For each $1 \le i \le n$ and for each $1 \le j \le m$, an entry $state[i, j]$ stores the state $EL_{min}(\{cs_x[0:i], cs_y[0:j]\})$, which means the minimal encoding length of $cs_x[0:i]$ and $cs_y[0:j]$. We fill out the entries $state[i, j]$ from smaller to larger values of $i + j$, so that $state[i, j]$ is computed after $state[i-1, j]$, $state[i, j-1]$, $state[i-1, j-1]$, …, $state[0, 0]$.

*Initialization*: For $i = 0$, we set all entries $state[0, j] = f^*(\{cs_y[0:j]\})$, where $f^*$ is the optimal encoding function for $C_y[0:j]$, and for $j = 0$, we set all entries $state[i, 0] = f^*(\{cs_x[0:i]\})$, where $f^*$ is the optimal encoding function for $cs_x[0:i]$.

Due to the objective of finding the best encoders for each residual subsequence and the varying properties of different encoders, for state transition, we need to traverse all previous states to compute the current state.

*State transition*: For $i, j > 0, state[i, j] = \min\{state[i-k, j-l] + EL_{min}(cs_x[i-k:i], cs_y[j-l:j])$, $k \in [0, i], l \in [0, j], k + l \ne 0\}$. The minimal encoding length is $state[n, m]$.

**Correctness.** To prove the correctness of the above dynamic programming algorithm, we show that for each $1 \le i \le n$ and for each $1 \le j \le m$, $state[i, j] = EL_{min}(cs_x[i], cs_y[j])$.

*Proof:* The proof is established through induction on the algorithm execution. The base case occurs when $i = 0$ or $j = 0$, at which point $state[i, j] = 0$, which is evidently correct. Now, consider an arbitrary entry $state[i, j]$ and assume that the claim holds for all entries $state[i', j']$ where $i' + j' < i + j$. Let $X' = EL_{min}(cs_x[i], cs_y[j])$. Suppose there exists a $state'[k, l]$ that is not the minimal encoding length increment for $cs_x[0:k]$ and $cs_y[0:l]$ ($state[k, l] < state'[k, l]$), but satisfies $X' = state'[k, l] + EL_{min}(cs_x[k:i], cs_y[l:j]) < state[k, l] + EL_{min}(cs_x[k:i], cs_y[l:j])$, where $k \in [0, i], l \in [0, j]$. Given that the value of $EL_{min}(cs_x[k:i], cs_y[l:j])$ is unique (i.e., the optimal encoding of $cs_x[k:i], cs_y[l:j]$ is unique), this contradicts $state[k, l] < state'[k, l]$. Therefore, the correctness of the algorithm holds.

**Time Complexity.** In accordance with the dynamic programming algorithm, filling each entry $state[i, j]$ requires traversing all entries from smaller to larger values of $i + j$. Assume that the two clusters consist of $N$ and $M$ records, respectively. To fill each entry $state[i, j]$ and traverse each entry from smaller to larger values of $i + j$, it is also necessary to enumerate all field encoding options and select the best one. Considering there are $n \times m$ entries in total, the time complexity of minimal encoding length merging is $O(|F| \times (N + M) \times n^2 \times m^2)$, where $|F|$ represents the number of encoding functions.

**Observation.** Due to the high computational complexity of the above algorithm, solving the minimal encoding length merging problem for arbitrary encoders in practice is infeasible. This is primarily because the transition from the previous to the current state costs $O(n \times m)$ time. Nonetheless, we observe that most commonly used encoding functions (e.g., VARCHAR) have the property that an increase in data length always results in an increase in encoding length. In other words, such encoding functions are *monotonic*. This observation allows us to reduce the search space from $n \times m$ to 3 (i.e., the most recent three states), significantly accelerating computation. We now formally define the monotonicity of encoding functions and illustrate the benefits with a simple example.

*Definition 4.* **Monotonicity.** *An encoding function set $F = \{f_1, f_2, ..., f_{|F|}\}$ is monotonic if it satisfies: For an arbitrary string $s$, $\forall i, j \in [1, |F|]$, $f_i(s[0:x]) \le f_j(s[0:x+1])$. An encoding function $f$ is monotonic if it satisfies: For an arbitrary string $s$, $f(s[0:x]) \le f(s[0:x+1])$.*

**Example 3.** In Figure 4, to compute the value at Ex.C, if the encoding function is monotonic, we only need to traverse the 3 blue entries instead of all 24 entries from its upper-left.

To exploit the monotonicity, we first define the following problem and then propose an improved dynamic programming algorithm.

**Problem 3** (Minimal Encoding Length Merging for Monotonicity Encoders). *Given a monotonic encoding function set $F = \{f_1, f_2, ..., f_{|F|}\}$, $N$ clusters $c_1, c_2, ..., c_N$, we define the merging of $c_i$ and $c_j$, $i \ne j$ and $i, j \in [1, N]$ as the minimal encoding length merging if the encoding length increment of merging $c_i$ and $c_j$ is minimal among all combinations of cluster pairs.*

To explain how to solve the minimal encoding length merging problem for monotonic encoders in practice, we take VARCHAR encoding as an example. First, given a string $s$ and its corresponding pattern $p$ we define the VARCHAR encoder as:

$$f_{vc}(s, p) = \{p_{id}, (l_1, s_1), (l_2, s_2), ..., (l_n, s_n)\}, \tag{2}$$

where $f_{vc}$ is the encoding function of VARCHAR. $p_{id}$ is the unique identity of pattern $p$, which usually takes up four bytes for a 32-bit unsigned integer. $s_1, s_2, \ldots, s_n$ are the residual subsequences of $s$ extracted by $p$, and $l_i$ ($i \in [1, n]$) is used to mark the length of $s_i$, which only takes up one or two bytes using the VARINT encoder. According to the given encoding function $f_{vc}$, the encoding length of a string is positively related to its length under VARCHAR. Clearly, it satisfies the monotonicity in Definition 4.

**Algorithm Description.** Similar to the previous algorithm, we need to fill a dynamic programming table with $n \times m$ entries, starting from the smaller to larger value of $i + j$. The *Initialization* remains the same as before. The only difference is in the transition:

*State transition*: For $i, j > 0$, if $cs_x[i] = cs_y[j]$ and both are not a wildcard, then $state[i][j]$ transitions from $state[i-1][j]$, $state[i][j-1]$, or $state[i-1][j-1]$, depending on the minimum value of the three. The minimal encoding length increment is $state[n, m]$. In detail, we use Algorithm 2 to compute the state transitions to $state[i][j]$. It uses 5 parameters: $cur\_state$ represents the current state that we need to compute. $type$ records the type (a pattern or a residual subsequence) of the source state (i.e., $state[i-1][j]$ or $state[i][j-1]$). For example, assume the source state is $state[i-1][j]$, then the type will be $isPattern$ if $cs_x[i-1]$ and $cs_y[j]$ are part of the pattern, and vice versa. $new\_char$ records the new character that we need to consider. For instance, if the state transitions from $state[i-1][j]$ (or $state[i][j-1]$), then $new\_char$ should be $C_x[i]$ (or $C_y[j]$). $size_x$ and $size_y$ represent the sizes of clusters $C_x$ and $C_y$, respectively.

First, if the previous state type is a pattern (i.e., $type = isPattern$), it requires an additional ($size_x + size_y$) length of descriptors to record the encoding length of new residual subsequences

**Algorithm 1** MinEncodingLengthIncrement($C_x$, $C_y$, $size_x$, $size_y$)

1: Let $cs_x$ ($cs_y$) be the common subsequence of the optimal pattern of $C_x$ ($C_y$);
2: $n \leftarrow$ the length of $cs_x$; $m \leftarrow$ the length of $cs_y$;
3: let $state[0...n][0...m]$ and $type[0...n][0...m]$ be new tables;
4: $state[0][0] \leftarrow 0$;
5: $type[0][0] \leftarrow isPattern$;
6: **for** $i \leftarrow 1$ to $n$ **do**
7:     $type[i][0] \leftarrow isRS$;
8:     $state[i][0] \leftarrow$ **UpdateState**($state[i-1][0], type[i-1][0], cs_x[i], size_x, size_y$);
9: **for** $j \leftarrow 1$ to $m$ **do**
10:     $type[0][j] \leftarrow isRS$;
11:     $state[0][j] \leftarrow$ **UpdateState**($state[0][j-1], type[0][j-1], cs_y[j], size_y, size_x$);
12: **for** $i \leftarrow 1$ to $n$ **do**
13:     **for** $j \leftarrow 1$ to $m$ **do**
14:         **if** $cs_x[i] = cs_y[j]$ and $cs_x[i] \neq *$ and $cs_y[j] \neq *$ **then**
15:             $state[i][j] \leftarrow \min(state[i-1][j-1],$ **UpdateState**($state[i-1][j], type[i-1][j], cs_x[i], size_x, size_y$), **UpdateState**($state[i][j-1], type[i][j-1], cs_y[j], size_y, size_x$));
16:             **if** $state[i-1][j-1]$ is the only minimal state **then**
17:                 $type[i][j] \leftarrow isRS$;
18:             **else**
19:                 $type[i][j] \leftarrow isPattern$;
20:         **else**
21:             $state[i][j] \leftarrow \min($**UpdateState**($state[i-1][j], type[i-1][j], cs_x[i], size_x, size_y$), **UpdateState**($state[i][j-1], type[i][j-1], cs_y[j], size_y, size_x$));
22:             $type[i][j] \leftarrow isRS$;
23: **return** $state[n][m]$;

**Algorithm 2** UpdateState($cur\_state$, $type$, $new\_char$, $size_x$, $size_y$)

1: **if** $type = isPattern$ **then**
2:     $cur\_state \leftarrow cur\_state + size_x + size_y$;
3: **if** $new\_char \neq *$ **then**
4:     $cur\_state \leftarrow cur\_state + size_x$;
5: **else**
6:     $cur\_state \leftarrow cur\_state - size_x$;
7: **return** $cur\_state$;

(lines 1-2). We then need to determine whether $new\_char$ is a wildcard $*$ or not. If not, $new\_char$ should be part of a residual subsequence. In this case, the encoding length should be increased by $size_x$ (lines 3-4), which is the number of $new\_char$ (the size of $C_x$). Otherwise, we need to subtract the extra $size_x$ length descriptors, which have already been counted in the encoding length when $C_x$ was computed (lines 5-6). For simplicity, we only discuss the computation of the minimal encoding length increment, but the computation of the corresponding optimal pattern is similar.

**Correctness.** We prove the correctness by showing that for each $1 \leq i \leq n$ and for each $1 \leq j \leq m$, $state[i, j] = EL_{min}(cs_x[i], cs_y[j])$.

*Proof:* The proof is by induction on the algorithm's execution. The base case is when $i = 0$ or $j = 0$. Then $state[i, j] = 0$, which is clearly correct. Consider now an entry $state[i, j]$ and assume that the claim holds for all entries $state[i', j']$ with $i' + j' < i + j$. Let $X' = EL_{min}\{cs_x[i], cs_y[j]\}$.

We discuss the following three cases: (1) $state[i, j]$ transitions from $state[i-1, j-1]$. (2) $state[i, j]$ transitions from $state[i-1, j]$. (3) $state[i, j]$ transitions from $state[i, j-1]$. Assume there is a $state'[k, l]$, which is not the minimal encode length increment for $A[0:i]$ and $B[0:j]$ (i.e., $state[k, l] < state'[k, l]$), but satisfied that $X' = state'[k, l] + EL'_{min}(cs_x[k:i], cs_y[l:j]) < state[k, l] + EL_{min}(cs_x[k:i], cs_y[l:j])$, $k \in [1, i], l \in [1, j]$.

For case (1), it is clearly that $EL_{min}(cs_x[k:i], cs_y[l:j]) = EL'_{min}(cs_x[k:i], cs_y[l:j]) = 0$, and it contradicts $state[k, l] < state'[k, l]$.

For case (2), the difference between $state[k, l]$ and $state'[k, l]$ is the multiple of $(size_x + size_y)$, due to different patterns generated by $cs_x[0:k]$ and $cs_y[0:l]$, i.e., $state[k, l] + size_x + size_y \leq state'[k, l]$. Considering lines 1-2 in Algorithm 2, the state of $EL_{min}(cs_x[k:i], cs_y[l:j])$ is $(size_x + size_y)$ more than that of $EL'_{min}(cs_x[k:i], cs_y[l:j])$ in the worse case, i.e., $EL_{min}(cs_x[k:i], cs_y[l:j]) \leq EL'_{min}(cs_x[k:i], cs_y[l:j]) + size_x + size_y$. Since our encoding function satisfies the monotonicity in Definition 4, it is obvious that $state'[k, l] + EL'_{min}(cs_x[k:i], cs_y[l:j]) \geq state[k, l] + EL_{min}(cs_x[k:i], cs_y[l:j])$ and it contradicts the assumption.

The proof of case (3) is similar to case (2). Thus the correctness holds.

**Complexity.** There are $n \times m$ entries in the $state$ table. To fill each value, we need to look up 3 previous states. Thus, the time complexity of Algorithm 1 is $O(n \times m)$, significantly reduced from $O(|F| \times (N + M) \times n^2 \times m^2)$, the complexity with an non-monotonic encoding function set $F$. Assume there are a string set $S$ initially, then it takes at most $|S| - k$ iterations of merging to finish the pattern extraction. In each iteration, it takes $|S| \times n \times m$ to update the pair-wise distance. Thus, the overall time complexity is $O(|S|^2 \times n \times m)$.

## 5 FURTHER OPTIMIZATIONS

### 5.1 Optimization of Pattern Extraction

**1-gram Distance Pruning.** Algorithm 1 repeats $N$ times in each iteration, because we need to compute the pairwise distance between the newly merged cluster and all the other clusters. To accelerate this process and avoid unnecessary computations, we propose a pruning strategy based on the *1-gram distance*.

*Definition 5.* **1-gram distance.** Given two strings $str_1$ and $str_2$, we build two multisets $MS_1$ and $MS_2$ for $str_1$ and $str_2$, respectively. Each element in the multisets is a symbol in the string. The 1-gram distance of $str_1$ and $str_2$ is:

$$Dist_1(str_1, str_2) = |MS_1 \cup MS_2| - 2 \times |MS_1 \cap MS_2|. \tag{3}$$

The *1-gram distance* can be easily extended to compute the distance between clusters with $O(n + m)$ time complexity. More importantly, the *1-gram distance* is a lower bound of the result of Algorithm 1. Thus, we can use it to prune the computation after each merging: Assume there are $N$ clusters $c_1, c_2, \ldots, c_N$ and $c_1$ is the newly merged one. To compute $EL_{min}$ between $c_1$ and $c_2, \ldots, c_N$, we maintain a lower bound value $MIN$ for it:

(1) Using Algorithm 1 to compute $EL_{min}(c_1, c_2)$ and use it to initialize $MIN$.

(2) For the subsequent computation ($EL_{min}(c_1, c_3), EL_{min}(c_1, c_3), \ldots, EL_{min}(c_1, c_N)$), we first compute the 1-gram distance. If the 1-gram distance is larger than $MIN$, then we skip the computation. Otherwise, we conduct Algorithm 1 to compute the exact encoding length increment.

(3) In the process of Algorithm 1 to compute the exact encoding length increment, if we find a state whose value is bigger than $MIN$, the computation would be terminated and skipped.

Using the above pruning strategy, we can replace the computation of Algorithm 1 with $O(n \times m)$ complexity to the computation of *1-gram distance* with $O(n + m)$ complexity.

## 5.2 Optimization of Compression Ratio

As we introduced in Section 4, it is challenging to determine the optimal clustering result and the optimal encoding for residual subsequences of each cluster in practice. Therefore, we can only consider the monotonic encoders (e.g., VARCHAR) in the pattern extraction process, which still has storage redundancy. To further reduce the redundancy of residual subsequences, we provide two options: (1) Using entropy encoding techniques (e.g., Huffman coding) or random access compression techniques (e.g., FSST) to encode the residual subsequences, which compresses strings one by one. (2) Using dictionary compression methods (e.g., Zstd and LZMA) to compress the residual subsequences, which is a block-wise compression.

## 6 THEORETICAL ANALYSIS

To quantify data redundancy and measure how clustering improves it, we consider the compression of each string as being generated by a joint distribution of two random variables: one representing the pattern and the other representing the characters in the residual subsequence. This distribution can be modeled by entropy [49], a traditional measure of randomness and uncertainty in random variables typically used to quantify data redundancy. We first analyze the entropy of a single string, and then show the equivalence between minimal encoding length and minimal entropy.

According to Section 3, each string can be divided into two parts: a pattern $p$ and a residual subsequence $r$. Consequently, the entropy (i.e., the average encoding length) of a string also consists of these two parts, which are given by:

$$E(S) = E(P) + |\overline{R}|E(R). \tag{4}$$

Equation 4 comprises two terms: $E(P)$ and $E(R)$. Each is weighted by the frequency with which it occurs in a single string. Specifically, $E(P)$ is weighted by 1 because it occurs only once in a string, and $E(R)$ is weighted by $|\overline{R}|$ because it occurs $|\overline{R}|$ times. $E(P)$ is the average entropy of a single pattern, a measure of uncertainty arising from assigning pattern labels to strings.

$$E(P) = \sum_{i=0}^{k} P_i \log P_i, \tag{5}$$

where $P_i$ represents the percentage of the $i$-th pattern. $E(R)$ is the average entropy of a single character in a residual subsequence, and $|\overline{R}|$ is the average length of residual subsequences.

$$E(R) = \sum_{i=0}^{k} P_i E(R_i), \tag{6}$$

where $E(R_i) = \sum_{j=0}^{D_i} p_{i,j} \log p_{i,j}$ ($p_{i,j}$ represents the frequency of occurrence of character $j$ in the $i$-th cluster) is the entropy of residual subsequences in strings with the $i$-th pattern, and is a measure of uncertainty arising from the distribution of characters within strings with the $i$-th pattern.

Since the entropy of a well-organized system should be low, the result of Problem 1 should have the overall entropy as low as possible. Inspired by [33], we model our compression from the entropy perspective as the following entropy-based optimization problem:

PROBLEM 4 (MINIMAL ENTROPY CLUSTERING). *Given encoding function $f$, a set of strings $S$, and a pattern size constraint $k$, the minimal entropy clustering problem is to find an optimal clustering of $S \rightarrow \{S_1, S_2, ..., S_k\}$ and corresponding optimal patterns $P = \{p_1, p_2, ..., p_k\}$ such that $E(S)$ is minimal.*

Following the proposed greedy algorithm to solve the clustering problem, we initially treat each string as a string set with the same pattern. Then, in each iteration, we use the entropy as a criterion, i.e., merge the two sets of strings that give rise to the largest decrease or the smallest increase in entropy.

Assuming that we are in an intermediate process of merging patterns, in the next iteration, we want to merge the two sets of strings $S_i$ and $S_j$. The discriminant function for merging $S_i$ and $S_j$ may be written as:

$$g(S_i, S_j) = E'(S) - E(S) \\ = (E'(P) + L'E'(R)) - (E(P) + LE(R)), \quad (7)$$

where $E'(S)$, $E'(P)$, and $E'(R)$ represent the updated $E(S)$, $E(P)$, and $E(R)$ after the merging, respectively.

Thus, the merging of patterns in each iteration is to find the two sets of strings that can minimize $g(S_i, S_j)$. Since $E(P)$ and $E'(P)$ are independent of $S_i, S_j$, the discriminant becomes:

$$g(S_i, S_j) = (L'E'(R) - LE(R)) \\ = (L' - L)E'(R) + L(E'(R) - E(R)), \quad (8)$$

where $L$ is the average length of residual subsequences and $L'$ is the updated $L$. As we assume that residual subsequences are generated by the random variable, the mathematical expectation of $E'(R) - E(R)$ is 0. The discriminant becomes:

$$g(S_i, S_j) = (L' - L), \quad (9)$$

where $L$ and $L'$ are the overall occurrences of symbols in residual subsequences of $S_i$ and $S_j$ before and after merging, respectively. When considering a specific encoding function of symbols, $L$ is $EL(S_i, p_i, f) + EL(S_j, p_j, f)$ and $L'$ is $EL(S_i \cup S_j, p, f)$, which implies that the merging of two sets for minimal encoding length increment (Problem 2) is equivalent to the merging of two sets for minimal entropy.

**Discussion.** As previously discussed, the minimization of encoding length is, to some extent, equivalent to the minimization of overall entropy. It reveals the reason why the proposed method works in data compression. However, there are some differences between these two criteria with different implicit assumptions. For Problem 4 (*Minimal Entropy Clustering*), we assume that all correlations among strings are extracted by finding patterns, and the remaining symbols in residual subsequences are randomly distributed. No further regularities can be exploited and used to compress the string. For Problem 1 (*Minimal Encoding Length Clustering*), we only consider the VARCHAR encoding and ignore the changes in character distribution and cardinality. Additionally, we assume the size of the pattern is fixed (user-defined according to the buffer size), and thus, we do not consider the storage cost of encoding patterns in the process of iterative merging. Instead, we only count the size of encoded residual subsequences.

## 7 EXPERIMENTAL EVALUATION

### 7.1 Experimental Setting

*7.1.1 Datasets.* We use two categories of datasets.

**Production Key-value Datasets.** There are a significant number of workloads in our in-memory key-value database, TierBase. The key-value data is generated by different applications according to real-world business requirements, making them perfect representatives of real-world machine-generated data. Specifically, we use five key-value datasets named from $KV1$ to $KV5$ for different use cases.

Table 2. Dataset statistics.

| Datasets | KV1 | KV2 | KV3 | KV4 | KV5 | Android | Apache | BGL | HDFS | Hadoop | AliLogs | github | cities | unece | urls | uuid |
|---|---|---|---|---|---|---|---|---|---|---|---|---|---|---|---|---|
| Num. of Records | 33.1B | 20.9B | 2.86M | 418K | 2.68M | 1.55M | 56.5K | 4.75M | 11.2M | 2.61M | 350K | 8.6K | 148K | 0.81K | 100K | 100K |
| Avg. Record Len. | 71.5 | 158.6 | 90.6 | 44.1 | 53.1 | 129.7 | 63.9 | 164.1 | 141.2 | 266.9 | 299.2 | 863.8 | 232.2 | 4494.8 | 63.1 | 35.6 |

**Public Datasets.** We choose two representative machine-generated datasets, log and JSON data. We use log datasets from different systems, including Android, Apache, HDFS, BGL, Hadoop, and an industrial Cloud [25, 58]. The JSON dataset *github* is curated from the test data of Zstd [11]. The *unece* and *cities* are both JSON datasets that record information about UNECE's countries and different cities in the world, respectively. We also curate URL dataset *urls* and UUID dataset *uuid* from [6], which are not typically machine-generated data but are generated by machines and can help us to determine the capacity boundaries of the proposed compression paradigm.

Table 2 demonstrates the details of the datasets.

*7.1.2* ***Environments.*** All the algorithms are implemented in C++ and compiled by Clang 13.0.1 with -O3 enabled. All experiments are performed on a machine with an Intel(R) Xeon(R) Platinum 8369B CPU @ 2.90GHz and 755GB main memory running Linux kernel version 4.19.91.

*7.1.3* ***Competitors.*** We study the following well-known compression algorithms in comparison with our algorithms.

- LZ (Lempel–Ziv) methods. In practice, existing NoSQL databases usually adopt LZ (Lempel–Ziv) methods for compression. As we surveyed in Section 2, Facebook RocksDB and Google LevelDB use Zstandard [11] and Snappy [24] to compress data, respectively. According to the benchmark results from [28, 52], we select the following algorithms as baselines:
  – *Zstandard (Zstd)* [11]: A compression method with a high compression ratio adopted by Facebook RocksDB [18].
  – *Snappy* [24]: A compression method adopted by Google *LevelDB* [23].
  – *LZ4* [64]: The best lightweight compression methods.
  – *LZMA* [61]: The compression method with the highest compression ratio in the LZ family.
- *FSST* [6]: A state-of-the-art general-purpose lightweight compression method which supports line-by-line compression.
- *LogReducer* [58]: A state-of-the-art log compression method.
- *Amazon Ion* [1]: A space-efficient JSON alternative from *Amazon*.
- *JSON BinPack* [54]: A state-of-the-art JSON-compatible serialization library that is designed for space-efficiency.
- The proposed *PBC* and its variants:
  – *PBC*: Our solution without further encoding for residual subsequences, which enables it to compress each data record individually, thus directly supporting random access to data.
  – $PBC_F$: Our solution with the *FSST* encoder, which can also compress each data record individually.
  – $PBC_Z$ & $PBC_L$: Our solutions with the *Zstd* encoder and *LZMA* encoder, respectively, which further improve the compression ratio on data blocks.

## 7.2 Effectiveness and Efficiency

We evaluate *PBC* with other competitors for the following measures: compression ratio, compression speed, and decompression speed. Specifically, the compression ratio is the ratio of the data size after compression to the raw data size.

Table 3. Line-by-line compression performance.

| Datssets | Comp. Ratio | | | | | Comp. Speed (MB/s) | | | | | Decomp. Speed (MB/s) | | | | |
|---|---|---|---|---|---|---|---|---|---|---|---|---|---|---|---|
| | FSST | LZ4 | Zstd | PBC | $PBC_F$ | FSST | LZ4 | Zstd | PBC | $PBC_F$ | FSST | LZ4 | Zstd | PBC | $PBC_F$ |
| KV1 | 0.393 | 0.504 | 0.577 | 0.236 | **0.147** | **237.64** | 23.29 | 11.98 | 33.85 | 32.95 | 2034.23 | 1243.25 | 256.59 | **2388.91** | 1330.62 |
| KV2 | 0.486 | 0.490 | 0.433 | 0.284 | **0.185** | **259.02** | 37.67 | 23.96 | 36.96 | 35.68 | 2525.01 | 1425.14 | 329.76 | **2536.24** | 1524.40 |
| KV3 | 0.307 | 0.371 | 0.423 | 0.239 | **0.134** | **308.32** | 30.43 | 16.01 | 66.36 | 63.30 | **2718.44** | 1831.32 | 351.82 | 2533.43 | 1490.73 |
| KV4 | 0.455 | 0.594 | 0.771 | 0.346 | **0.215** | **166.84** | 15.35 | 8.02 | 45.03 | 42.36 | 1540.76 | 1120.26 | 199.83 | **1753.95** | 967.12 |
| KV5 | 0.545 | 0.438 | 0.596 | 0.241 | **0.211** | **199.12** | 17.56 | 12.55 | 35.16 | 33.76 | 1496.42 | 1638.77 | 271.31 | **1920.56** | 1068.93 |
| Android | 0.576 | 0.560 | 0.543 | 0.347 | **0.245** | **261.78** | 30.92 | 18.62 | 60.40 | 53.13 | 2096.87 | 1282.84 | 284.01 | **3231.56** | 1580.05 |
| Apache | 0.322 | 0.349 | 0.411 | 0.151 | **0.104** | **320.72** | 31.31 | 12.07 | 48.85 | 43.32 | 3039.89 | 1773.38 | 343.56 | **3140.39** | 1909.66 |
| BGL | 0.293 | 0.376 | 0.356 | 0.325 | **0.146** | **391.88** | 42.68 | 23.97 | 34.64 | 30.71 | **3689.13** | 1528.33 | 344.38 | 1775.44 | 1314.49 |
| HDFS | 0.288 | 0.374 | 0.353 | 0.308 | **0.147** | **389.47** | 41.24 | 22.60 | 95.68 | 86.19 | **3591.41** | 2173.54 | 404.36 | 3135.02 | 1954.40 |
| Hadoop | 0.286 | 0.215 | 0.196 | 0.157 | **0.075** | **549.19** | 74.07 | 41.30 | 122.31 | 86.06 | 4464.45 | 3798.20 | 773.56 | **5412.87** | 3627.94 |
| AliLogs | 0.484 | 0.516 | 0.436 | 0.425 | **0.347** | **348.83** | 64.26 | 40.01 | 131.95 | 93.22 | 3155.18 | 2923.25 | 420.59 | **4623.95** | 2038.80 |
| cities | 0.316 | 0.336 | 0.305 | 0.261 | **0.189** | **392.78** | 58.21 | 34.86 | 58.44 | 54.82 | **3776.21** | 2114.45 | 506.31 | 2296.38 | 1550.50 |
| github | 0.278 | 0.151 | 0.101 | 0.110 | **0.092** | **467.30** | 161.87 | 102.57 | 65.92 | 62.18 | 5148.69 | 3975.35 | 1671.84 | **7433.55** | 4722.06 |
| unece | 0.437 | 0.210 | 0.125 | 0.106 | **0.057** | **303.25** | 189.72 | 156.80 | 89.15 | 79.17 | 3890.06 | 2579.46 | 1192.75 | **8418.96** | 6224.83 |
| urls | 0.413 | 0.456 | 0.611 | 0.299 | **0.248** | **195.89** | 22.15 | 11.35 | 63.67 | 55.11 | 1807.98 | 1247.63 | 158.91 | **2029.16** | 1043.43 |
| uuid | 0.443 | 0.788 | 0.984 | 0.721 | **0.421** | **149.32** | 13.61 | 6.76 | 117.39 | 82.78 | 1776.21 | **2158.41** | 166.14 | 1776.11 | 907.47 |

*7.2.1 **Line-by-Line Compression.*** Since there is a significant demand for storing relatively short data records individually for fast random access in database systems, we evaluate the competitors in the setting of compressing every data record individually, which naturally supports random access. According to [6], the online methods in the LZ family cannot handle short data records reasonably. Their compression ratio is near 1 or even over 1 in most datasets, which means that the data records are not effectively compressed. Since *LZ4* and *Zstd* can considerably improve the compression ratio for short records by using an additional dictionary trained on raw data, we evaluate *LZ4* and *Zstd* with a pre-trained dictionary (trained by *Zstd*) on the default compression level, denoted *LZ4(dict)* and *Zstd(dict)*.

Note that *LZ4(dict)*, *Zstd(dict)*, and our proposed *PBC* and its variant $PBC_F$ all benefit from an offline training phase. In the experiments, we exclude the offline training time from the compression time because both the *Zstd*/*LZ4* dictionary and the *PBC* patterns can be used as long as the data generation patterns in the given application do not change (e.g., no alterations to the data model), as we explained in Section 7.5. The offline training allows *PBC* to extract structural information without any prior knowledge, and this overhead does not impact online compression.

As shown in Table 3, *FSST* [6] achieves the best performance in compression speed and works well in decompression, showing that *FSST* is one of the best lightweight compression methods. However, its compression ratio is not competitive when compared to other methods. The compression ratios of *PBC* and $PBC_F$ are significantly better than all the baselines on the majority of datasets. At the same time, *PBC* and $PBC_F$ also provide an intermediate-level compression speed and a fast decompression speed. Especially *PBC*, which can provide the state-of-the-art decompression speed in most datasets, compresses data by reducing the reduplication of patterns without any other encoding. The result shows that *PBC* and $PBC_F$ can save 50% more storage space than all the baselines with adequate compression and decompression performance. Additionally, the considerable gap in compression ratio between $PBC_F$ and *FSST* justifies the idea that the proposed pattern-based compression schema is orthogonal to *FSST* and can significantly improve the compression ratio.

*7.2.2 **Random Access.*** As shown in Table 3, compressing data records individually with existing methods results in a relatively poor compression ratio. Therefore, existing database systems typically utilize general-purpose data compression libraries to compress data on data blocks, which cannot directly support random access but can achieve a much higher compression ratio. In existing key-value database systems with block-wise compression, to access a specific data record *s*, we must first find the data block containing record $s$ and then decompress the whole data block to look up the record $s$.

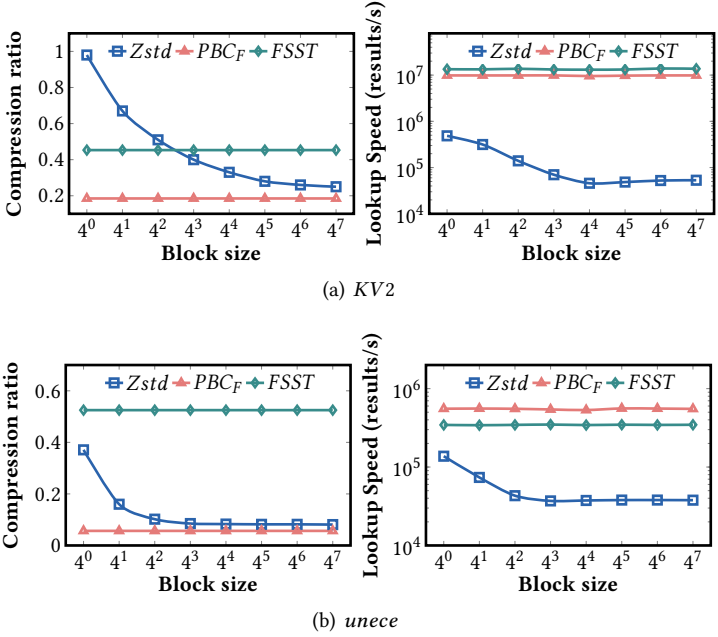

Fig. 5. Random access performance.

Both *RocksDB* [18] and our TierBase compress data with *Zstd* [11], which provides the best trade-off between compression ratio and efficiency for database systems. Thus, in Figure 5, we evaluate the performance of *Zstd*, *FSST*, and the proposed $PBC_F$ for randomly accessing data records. We compress data records into blocks (the block size is the number of records) and randomly sample 1% of data records to look up the compressed data. The compression ratio and lookup speed (results/s) on our production dataset $KV2$ and public JSON datasets *unece* are reported.

As shown in Figure 5, with the increase of block size, the compression ratio of *Zstd* improves to a considerable level, but the lookup speed deteriorates. On the contrary, $PBC_F$ and *FSST* are unaffected in compression ratio and still provide a superior lookup speed. Based on this, $PBC_F$ can also provide the best compression ratio, which makes $PBC_F$ much more attractive for database use cases.

7.2.3 **File Compression.** In this experiment, we concatenate all data records in the same dataset as a whole file. This is the most suitable case for Lempel-Ziv (LZ) baselines because it has large blocks to compress. Although *PBC* can naturally compress each data individually, due to the pattern-based schema, *PBC* is orthogonal to most existing compression methods and can cooperate with them to improve the compression ratio. It can work in conjunction with block compression methods to provide a better compression ratio than *PBC* and $PBC_F$. We evaluate the performance of four representative compression methods and two *PBC* variants. Specifically, $PBC_Z$ and $PBC_L$ utilize *Zstd* and *LZMA* as the encoding backbone after extracting patterns, respectively. We measure the compression ratio, bulk compression speed, and bulk decompression speed.

As shown in Table 4, $PBC_L$ and $PBC_Z$ can provide the best compression ratio on most datasets. Note that the worst compression ratio superiority occurs in the uuid dataset, which is more like random-generated data than machine-generated data, thus having fewer structural characteristics.

Table 4. File compression performance.

| Datasets | Comp. Ratio | | | | | | Comp. Speed (MB/s) | | | | | | Decomp. Speed (MB/s) | | | | | |
| --- | --- | --- | --- | --- | --- | --- | --- | --- | --- | --- | --- | --- | --- | --- | --- | --- | --- | --- |
| | Snappy | LZMA | LZ4 | Zstd | $PBC_Z$ | $PBC_L$ | Snappy | LZMA | LZ4 | Zstd | $PBC_Z$ | $PBC_L$ | Snappy | LZMA | LZ4 | Zstd | $PBC_Z$ | $PBC_L$ |
| KV1 | 0.345 | 0.138 | 0.339 | 0.192 | 0.133 | **0.109** | 442.12 | 2.67 | **644.55** | 302.40 | 31.85 | 3.82 | 1624.23 | 143.23 | **3116.07** | 795.62 | 1138.56 | 216.21 |
| KV2 | 0.449 | 0.131 | 0.436 | 0.209 | 0.142 | **0.100** | 358.40 | 2.72 | **482.90** | 285.61 | 34.82 | 3.77 | 1338.32 | 131.42 | **3164.86** | 762.93 | 1291.86 | 183.08 |
| KV3 | 0.243 | 0.109 | 0.233 | 0.140 | 0.106 | **0.080** | 666.19 | 2.99 | **871.50** | 400.71 | 60.59 | 5.00 | 2278.10 | 176.07 | **3582.55** | 962.41 | 1333.60 | 229.13 |
| KV4 | 0.427 | 0.183 | 0.435 | 0.255 | 0.192 | **0.161** | 371.18 | 2.70 | **555.85** | 245.25 | 40.09 | 5.42 | 1391.56 | 102.03 | **3364.69** | 796.68 | 809.83 | 138.49 |
| KV5 | 0.229 | 0.078 | 0.182 | 0.102 | 0.090 | **0.066** | 755.51 | 5.19 | **1026.92** | 577.65 | 33.89 | 7.31 | 2672.19 | 246.94 | **4587.84** | 1548.87 | 1159.69 | 257.61 |
| Android | 0.232 | 0.053 | 0.197 | 0.078 | 0.059 | **0.038** | 645.47 | 8.97 | **879.05** | 662.95 | 54.34 | 9.10 | 2516.78 | 350.99 | **3554.06** | 1436.71 | 1347.37 | 403.69 |
| Apache | 0.108 | 0.040 | 0.088 | 0.053 | 0.038 | **0.027** | 1490.88 | 13.18 | **1933.66** | 1139.18 | 45.52 | 20.13 | 5373.96 | 440.58 | **5569.54** | 2004.86 | 1769.31 | 516.63 |
| BGL | 0.169 | 0.057 | 0.167 | 0.094 | 0.080 | **0.041** | 1044.26 | 6.53 | **1228.31** | 572.92 | 31.53 | 5.36 | 3317.08 | 289.46 | **4271.77** | 1166.69 | 851.35 | 300.35 |
| HDFS | 0.182 | 0.074 | 0.176 | 0.096 | 0.072 | **0.051** | 869.65 | 6.22 | **1022.50** | 600.72 | 85.66 | 5.85 | 3150.80 | 219.29 | **4039.20** | 1191.26 | 1311.86 | 331.55 |
| Hadoop | 0.108 | 0.044 | 0.086 | 0.048 | 0.038 | **0.023** | 1794.74 | 12.30 | **2076.36** | 1021.57 | 112.84 | 13.00 | 4732.46 | 427.18 | **6563.69** | 2056.79 | 2254.04 | 610.91 |
| AliLogs | 0.463 | 0.288 | 0.456 | 0.312 | 0.279 | **0.265** | 928.36 | 8.21 | **1096.45** | 364.97 | 92.59 | 12.03 | 3164.64 | 79.07 | **5953.70** | 1183.83 | 1484.44 | 91.27 |
| cities | 0.205 | 0.077 | 0.172 | 0.120 | 0.099 | **0.075** | 850.89 | 4.95 | **1074.56** | 472.33 | 54.36 | 10.04 | 2953.54 | 211.24 | **5927.21** | 1352.51 | 1250.01 | 209.94 |
| github | 0.103 | 0.055 | 0.117 | 0.062 | 0.014 | **0.012** | 2163.92 | 12.29 | **2340.23** | 1081.47 | 64.91 | 44.90 | 5960.64 | 382.90 | **9768.09** | 2170.38 | 4493.18 | 1367.49 |
| unece | 0.201 | 0.069 | 0.172 | 0.090 | 0.049 | **0.042** | 787.03 | 8.35 | **1005.69** | 607.99 | 83.37 | 29.35 | 3162.26 | 273.57 | **4435.54** | 1600.10 | 3620.45 | 408.46 |
| urls | 0.361 | 0.151 | 0.355 | 0.208 | 0.158 | **0.122** | 473.28 | 4.52 | **642.47** | 299.13 | 56.58 | 9.37 | 1858.42 | 146.87 | **3172.54** | 960.62 | 1073.13 | 166.11 |
| uuid | 0.687 | 0.347 | 0.687 | 0.400 | 0.396 | **0.346** | 366.45 | 2.73 | **628.51** | 199.46 | 76.09 | 4.28 | 1418.09 | 50.84 | **4194.16** | 925.43 | 660.18 | 50.16 |

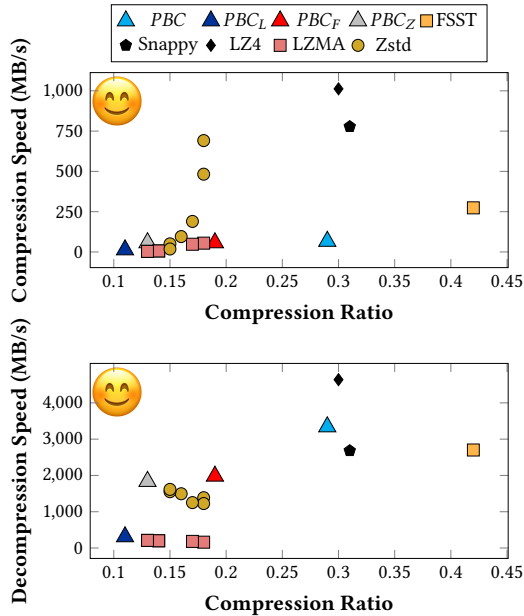

Fig. 6. Performance visualization of compared methods.

Compared with lightweight compression methods *Snappy* and *LZ4*, *PBC* variants may not provide similar efficiency but can offer a much better compression ratio. As for *Zstd* and *LZMA*, the cooperation with *PBC* influences compression and decompression speed. For *Zstd*, it harms the compression speed but improves decompression efficiency. For *LZMA*, it significantly improves both effectiveness and efficiency. The influences on compression/decompression ratio arise because the *PBC* process can considerably compress raw data (please refer to the result of *PBC* in Table 3), thus speeding up subsequent operations. The influences on compression ratio mainly stem from the fact that most existing methods compress data by encoding frequently repeated substrings, while *PBC* focuses on reduplicated subsequences, which allows *PBC* to capture redundancy differently from existing methods.

### 7.2.4 *Pareto-Optimal Compression for Machine-Generated Data.* To better illustrate the position of *PBC* for machine-generated data compression, we draw two figures depicting both

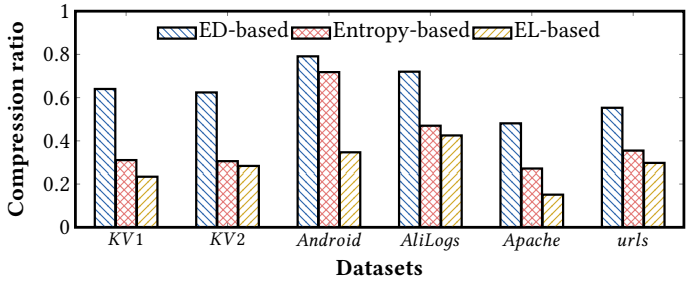

Fig. 7. Effect of clustering criteria.

compression and decompression performance based on the experimental results over all our datasets (top-left is better). Note that we plot *PBC* variants as triangles to distinguish them from baselines. There are multiple data points of *Zstd* and *LZMA* due to their multiple compression levels. As shown in Figure 6, when focusing on compression speed, *PBC*, $PBC_L$, *Zstd*, and *LZ4* are Pareto-optimal and constitute the Pareto frontier together. Moreover, in scenarios where decompression speed is more valued (e.g., read-intensive applications), *PBC* and its variants occupy 4 out of 5 positions in the Pareto frontier.

### 7.3 Ablation Study and Parameter Tuning

*7.3.1* **Effect of Clustering Criterion.** Based on the proposed compression schema, a naive solution is to use edit distance as the criterion for clustering. Moreover, in Section 6, we proposed the minimal entropy clustering problem. In this experiment, we compare the proposed encoding length-based (EL-based) criterion to the above two baselines (edit distance-based criterion and entropy-based criterion) in terms of compression ratio. As shown in Figure 7, the EL-based criterion and entropy-based criterion are both much better than the naive edit distance-based (ED-based) criterion, which demonstrates the effectiveness of the proposed techniques. The proposed EL-based criterion is also superior to the entropy-based criterion and achieves the best performance on all the datasets. This result indicates that although these two criteria both have some implicit assumptions (as discussed in Section 6), the proposed EL-based criterion is a better heuristic to guide clustering.

*7.3.2* **Efficiency of Pattern Extraction Optimization.** In this experiment, we provide the time cost of our pattern extraction phase (the extracted patterns are used in Section 7.2) with and without the *1-gram distance pruning* technique. As shown in Figure 8, the *1-gram distance pruning* can significantly reduce the running time of pattern extraction. Moreover, we observe that even on the largest dataset, $KV1$, the pattern extraction phase can be completed within an hour, which is an acceptable cost for an offline computation.

*7.3.3* **Effect of Sample Set Size.** To understand how the size of samples for pattern extraction affects the compression ratio, we report the compression ratio of *PBC* on two datasets under different sample sizes in Figure 9(a). In most datasets, when the sample size is over 2MB, the compression ratio converges to a sufficiently good value. It is important to note that we only need to sample a small proportion of data for pattern extraction. For example, for $KV1$ and $KV2$, we only sample 5.4MB and 4.5MB of data for extracting the patterns, respectively. *PBC* can achieve 23.6% and 20.9% compression ratios on 2.3TB of $KV1$ and 3.2TB of $KV2$, respectively.

*7.3.4* **Effect of Pattern Size.** To understand how the number of patterns affects performance, we report the compression ratio on two datasets under different pattern sizes in Figure 9(b). The compression ratio has a trend of decreasing with the reduction of pattern size. The law of diminishing marginal utility applies here, as the compression ratio can converge to a sufficiently good value when the pattern size is relatively large. In practice, we recommend users to set the pattern size according to the cache budget.

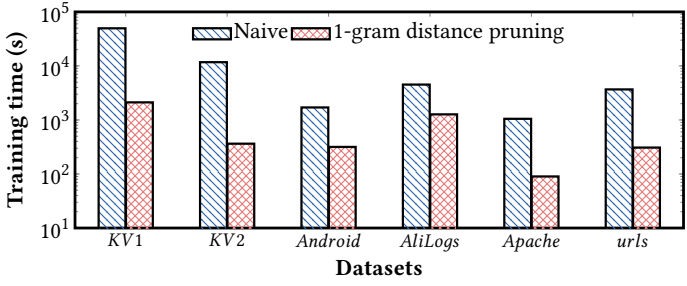

Fig. 8. Running time of pattern extraction.

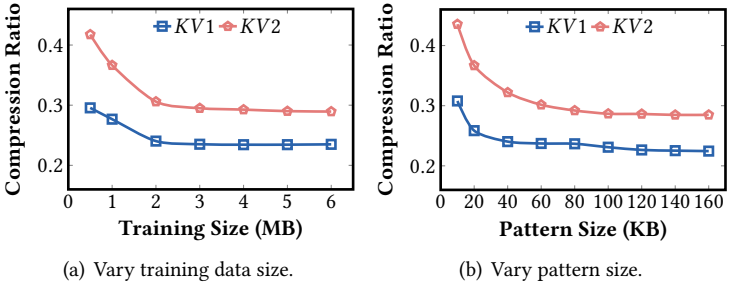

(a) Vary training data size.

(b) Vary pattern size.

Fig. 9. Effect of training data size and pattern size.

## 7.4 Performance on Specific Data Types

*7.4.1 Performance on Log Compression.* To further demonstrate the excellent performance of the proposed method, we compare a state-of-the-art log compression method, *LogReducer*, with $PBC_L$ (setting the compression level of *LZMA* at 9). *LogReducer* is a compression method specifically designed for log data compression. It relies on a log parser to identify structured data and proposes specific encoding techniques for timestamps and numerical variables. In other words, it compresses data with the aid of prior knowledge about data properties. Consequently, it is incapable of compressing other types of data, and we only compare it on log datasets. We report the average compression ratio and efficiency across all the log datasets in Table 2. As shown in Table 5, $PBC_L$ not only attains a similar compression ratio to *LogReducer* (owing to the timestamps and numerical variables encoding designed for log data, *LogReducer* is slightly better), but also achieves a much better performance in terms of compression and decompression speed.

*7.4.2 Performance on JSON Compression.* According to [50, 53], we also compare *PBC* with the best methods for JSON compression, *Amazon Ion* with the binary format and *JSON BinPack* with the schema-driven model, denoted as *Ion-B* and *BP-D*, respectively. Both methods are specifically designed for JSON data, and *BP-D* relies on an application-provided schema to identify the structure of a JSON document. In other words, they only work on JSON data and *BP-D* requires additional information about data formats and properties. We report the average compression ratio and efficiency across all our JSON datasets. As Table 6 shows, *PBC* and $PBC_F$ significantly outperform *Ion-B* and *BP-D* in compressing single records. For compressing an entire file (containing many JSON records), $PBC_L$ and BP-D both achieve excellent compression ratios. In Table 7, we provide a detailed comparison between the two best methods (BP-D+LZMA and $PBC_L$) in terms of compression ratio on different datasets below.

Table 5. Performance on log compression.

| Algorithms | Comp. Ratio | Comp. Speed | Decomp. Speed |
|---|---|---|---|
| *LogReducer* | 0.219 | 7.23 MB/s | 12.72 MB/s |
| $PBC_L$ | 0.224 | 13.8 MB/s | 169.5 MB/s |

Table 6. Performance on JSON compression.

|  | Record Compression | | | | File Compression | | |
|---|---|---|---|---|---|---|---|
| Methods | Ion-B | BP-D | PBC | $PBC_F$ | Ion-B +LZMA | BP-D +LZMA | $PBC_L$ |
| Comp. Ratio | 0.439 | 0.409 | 0.159 | **0.113** | 0.051 | **0.041** | 0.043 |
| Comp. Speed | 37.25 | 10.97 | **71.17** | 65.39 | 9.727 | 4.969 | **28.10** |
| Decomp. Speed | 55.07 | 23.73 | **6050** | 4166 | 44.27 | 22.53 | **662.0** |

Table 7. Performance on different JSON datasets.

| Dataset | *cities* | *github* | *unece* |
|---|---|---|---|
| BP-D | **0.072** | 0.029 | **0.023** |
| $PBC_L$ | 0.075 | **0.012** | 0.042 |

Owing to the ground-truth JSON schema provided by the application, the BP-D-based compression significantly outperforms $PBC_L$ on the *unece* dataset. Interestingly, $PBC_L$ still achieves a comparable compression ratio (and in some cases, such as the *github* dataset, significantly better).

An insight we gained from these results is that the optimal compression may not be achieved by the optimal semantic interpretation of the data. Some co-occurring relationships in the data may exist beyond semantic relationships. Taking JSON as an example, the optimal semantic interpretation (i.e., schema) can only capture co-occurrence at the key level, but not among values. This allows $PBC_L$ to perform better than JSON-specialized methods, underlining the motivation behind our proposed PBC.

### 7.5 Case Study in A Production Database System

A critical application of PBC is to compress data stored in production distributed in-memory NoSQL database systems (e.g., serialized data objects), which is typically generated by programs. In TierBase, our production database system, we have previously designed a dictionary-based compression solution. First, we sample data for a target workload and train a workload-specific dictionary using Zstd's training utility and a well-designed parameter search in an offline manner. Then, the dictionary is distributed to all instances of this workload. Subsequently, instances can utilize this dictionary to compress and decompress their data with Zstd. Furthermore, a monitoring component in TierBase continuously observes the overall compression ratio. If it deteriorates to a predefined threshold, a re-sampling and re-training process will be triggered.

Similarly, to integrate $PBC_F$ with TierBase, we sample data and extract patterns offline for each workload, and then use these patterns to compress data in all instances of the workload. In practice, unless the program of data generation in this workload (application) changes (e.g., a new version is released), these patterns will be effective for an extended period. To ensure that compression remains effective, we design a similar re-training mechanism: we set a counter to count the number of data records that do not match a pattern. Once the rate of unmatched records reaches a predefined threshold, a re-sampling and re-training program will be triggered.

Table 8.  Case study in a production database system.

| Metric | Memory Usage (%) | | Average SET Throughput($QPS$) | | Average GET Throughput($QPS$) | |
|---|---|---|---|---|---|---|
| Workload | A | B | A | B | A | B |
| Uncompressed | 100% | 100% | 125409 | 123167 | 131285 | 134807 |
| Zstd | 45% | 37% | 80914 | 91432 | 110302 | 130140 |
| $PBC_F$ | 25% | 29% | 84719 | 100088 | 129947 | 130893 |

After integrating $PBC_F$ as a compression option, we evaluate TierBase with $PBC_F$ compression and Zstd compression on two typical workloads. We measure the overall storage usage for all records (with the storage usage of uncompressed data set at 100%), and the throughput of both SET and GET commands (write and read data) for each single-threaded instance. As shown in Table 8, $PBC_F$ significantly improves storage usage, which can help us save valuable memory. Compared with Zstd, $PBC_F$ has better performance for both writing and reading data. In addition, the cost of pattern extraction for $PBC_F$ is similar to the cost of training a parameter-tuned dictionary for Zstd in TierBase, with both taking approximately 30 minutes for each workload.

## 8 CONCLUSION

We proposed Pattern-Based Compression (PBC), a high-ratio compression algorithm for compressing machine-generated data. It reduces data redundancy by exploiting representative common subsequences from data records. We modeled pattern-based compression as an optimal clustering problem and proposed the encoding length as a criterion for clustering. We designed an efficient algorithm to address the clustering problem and discover patterns. Theoretical analysis showed that the clustering guides the compression towards minimizing entropy. For line-by-line compression, PBC significantly outperformed state-of-the-art solutions, saving about half of the storage space compared to existing solutions. In contrast to block-wise compression methods, PBC demonstrated exceptional random access performance while maintaining a state-of-the-art compression ratio, making PBC particularly useful for storage services aiming to substantially save hardware costs with minimal overhead.